\documentclass[letterpaper, 10 pt, conference]{ieeeconf}

\usepackage{amsmath}

\let\labelindent\relax
\usepackage[T1]{fontenc}
\usepackage{amsthm}
\usepackage{amsfonts}
\usepackage{amssymb}  
\usepackage{mathtools}
\usepackage[utf8]{inputenc}
\usepackage{comment}
\usepackage{algorithm,algpseudocode}
\algblock{Input}{EndInput}
\algnotext{EndInput}
\algblock{Output}{EndOutput}
\algnotext{EndOutput}

\usepackage{breqn}
\usepackage{graphicx}
\usepackage[version=4]{mhchem}
\usepackage{siunitx}
\usepackage{longtable,tabularx}
\usepackage{graphics} 
\usepackage{float}
\usepackage{epsfig} 
\usepackage{times} 
\usepackage{color}
\usepackage[dvipsnames]{xcolor}
\usepackage{pifont}
\usepackage{enumitem}
\usepackage{bm}
\usepackage{bbm}
\usepackage{soul}
\usepackage{balance}
\makeatletter
\newcommand{\multiline}[1]{%
  \begin{tabularx}{\dimexpr\linewidth-\ALG@thistlm}[t]{@{}X@{}}
    #1
  \end{tabularx}
}
\makeatother
\usepackage{hyperref}
\hypersetup{
    colorlinks=true,
    linkcolor=blue,
    filecolor=blue,      
    urlcolor=blue,
    pdftitle={Overleaf Example},
    pdfpagemode=FullScreen,
    }

\newcommand{\trace}{\textit{tr}}

\newcommand{\E}{\mathbb{E\,}}

\newcommand{\R}{\mathbb{R}}

\newtheorem{assumption}{Assumption}

\newtheorem{remark}{Remark}

\newtheorem*{corollary}{Corollary}

\IEEEoverridecommandlockouts                              

\title{\Large \bf Dampening parameter distributional shifts under robust control and gain scheduling}
\author{Mohammad S. Ramadan,\,Mihai Anitescu
\thanks{The authors are with the Mathematics and Computer Science Division, Argonne National Laboratory, Lemont, IL 60439, USA,  {\tt\footnotesize mramadan@anl.gov, anitescu@mcs.anl.gov.}}
\thanks{This material was based on work
supported by the U.S. Department of Energy, Office of Science,
Office of Advanced Scientific Computing Research (ASCR) under
Contract DE-AC02-06CH11347. }
}

\begin{document}
\maketitle

\begin{abstract} 
Many traditional robust control approaches assume linearity of the system and independence between the system state-input and the parameters of its approximant (possibly lower-order) model. This assumption implies that the application of robust control design to the underlying system introduces no distributional shifts in the parameters of its approximant model. This is generally not true when the underlying system is nonlinear, which may require different approximant models with different parameter distributions when operated at different regions of the state-input space. Therefore, a robust controller has to be robust under the approximant model with parameter distribution that will be experienced in the future data, after applying this control, not the parameter distribution seen in the learning data or assumed in the design. In this paper, we seek a solution to this problem by restricting the newly designed closed-loop system to be consistent with the learning data and slowing down any distributional shifts in the state-input space of the underlying system, and therefore, in the parameter space of its approximant model. In computational terms, the objective of dampening the shifts in the parameter distribution is formulated as a convex semi-definite program that can be solved efficiently by standard software packages. We evaluate the proposed approach on a simple yet telling gain-scheduling problem, which can be equivalently posed as a robust control problem.
\end{abstract}

\section{Introduction} \label{section: Introduction}
Robust control and gain scheduling are important design concepts in the control of dynamic systems with uncertainties and nonlinearities \cite{petersen2014robust}, found in fields such as power systems \cite{mohammadi2021robust,rimorov2022gain}, robotics \cite{ryan2013lmi}, aerospace engineering \cite{liu2023robust} and many more. These control strategies typically rely on the principle of quadratic stability, which ensures system stability under varying operational conditions \cite{dorato1987historical}, rather than one condition as in the certainty equivalence case \cite{aastrom2012introduction}, or the local approximation about a reference point under the small signal model assumption. Traditional approaches to robust control and gain scheduling, however, often presume that a low-order model of a fixed parameter distribution is capable of capturing the behavior of the system (in the difference/differential inclusion sense) under any new control design \cite{scherer2006lmi}. This is generally problematic in nonlinear systems where the prediction accuracy of the approximant (and possibly lower-order) model with the current parameter distribution can be impaired under the application of a new control policy. This invalidates the premises for quadratic stability conditions used to guarantee the safety of these approaches.

In a previous work \cite{ramadan2024data}, we introduced a framework for control design that enforces the new closed-loop system to admit state-input distributions that are similar  (under some metric) to the distribution of the data used in the learning/identification step, thus addressing the premature generalization problem that arises when traditional control design methods falsely extrapolate the behavior of the system beyond the behavior captured by the data. We call a control design under this framework \textit{data-conforming}. The goal of this work is to show that the data-conforming framework is effective in solidifying the quadratic stability condition required for robust control and gain scheduling to retain their guarantees.

Mathematically, to equip the robust control and gain-scheduling methods with the data-conforming framework, we incorporate affine regularization terms and linear matrix inequality (LMI) constraints that enforce the consistency to the learning data. This formulation conserves the nature of convex semi-definite-program (SDP)-based robust control and gain-scheduling problems, offering a level of scalability to handle problems with high state-input dimensions and, hence, guarantees the applicability of our framework to various real-world problems.

The underlying philosophy of our design framework, although shares similarities to other frameworks in the field, prioritizes the standard control-theoretic formulations and preserves their computational efficiency. On a different front, enforcing consistency with the learning data is fundamental to offline reinforcement learning algorithms \cite{agarwal2020optimistic}. However, these algorithms are generally complex \cite{fujimoto2021minimalist} and rely on stochastic-gradient-based optimization, hindering their adaptation to robust control design formulations. The proposed robust data-conforming control design also shows similarities to the unfalsified control paradigm \cite{cabral2004unfalsified} in ensuring a degree of consistency with past data. Contrasting with this approach, which models consistency through a membership test, our approach is built around consistency in distances between distributions, allowing for generalized and statistics-based formulations capable of handling stochastic dynamics. Moreover, our approach is compatible with modern optimal and robust control design techniques by augmenting their standard formulations with affine regularization terms and convex constraints. We summarize our contributions with the following:
\begin{itemize}
    \item We explain how the application of robust control can itself invalidate robust control, by introducing parameter distributional shifts that weaken the quadratic stability condition required to achieve robustness in the first place.
    \item We adapt the data-conforming framework to the robust control and gain scheduling design concepts while preserving the computational efficiency and design practicality of these concepts.
    \item We present a simple yet telling example that shows how easy it is to invalidate the quadratic stability condition and result in poor robustness when standard robust control is used.
\end{itemize}

\section{Problem formulation} \label{section: Problem Formulation}
Consider the dynamic system
\begin{align}
x_{k+1}&=f(x_k, u_k, w_k), \label{eq:stateEquation}
\end{align}
where $x_k\in\mathbb R^{r_x}$ is the state, $u_k\in\mathbb R^{r_u}$ is the control input, $w_k \in \R^{r_x}$ is a white noise of zero mean and positive semi-definite covariance $W \succeq 0$, and $f$ is a locally bounded function in its arguments (differentiable in its first two arguments in the gain scheduling case).

The goal of this paper is to design a control law that minimizes the quadratic cost function (the steady-state weighted covariances of the state and input)
\begin{align}
    J = \lim_{k \to \infty} \E \left \{x_k^\top Q x_k + u_k^\top R u_k \right \}, \label{eq:costFunction}
\end{align}
where $Q \succeq 0$ and $R \succ 0$ are positive semi-definite and positive definite weighting matrices, respectively, and the expectation $\E$ is taken over all values of $w_k$ and $u_k$ (a stochastic policy will be used to ensure the persistence of excitation condition \cite{de2019formulas} for continual learning).

\begin{remark} \label{remark: 2 cases covered}
We target two possible scenarios:
\begin{enumerate}
    \item The dynamics $f$ is unknown, but state-input data are available and a data-driven control design is to be implemented.
    \item $f$ is known and differentiable in its first two arguments, and we seek to implement a gain-scheduling controller.
\end{enumerate}
\end{remark}

The following assumption is a modeling assumption that can cover both cases.

\begin{assumption} \label{assumption:diff inclusion}
Under some state-feedback control gain $K_0$, the behavior of the system can be modeled by the difference inclusion \cite{boyd1993control}
\begin{align} \label{eq:diff inclusion}
\begin{aligned}
    x_{k+1} &= F_k x_k +  G_k u_k, \\
    (F_k,G_k) &\in \mathbb C := \text{conv-hull}\left \{(A_i,B_i), i=1,\hdots,n \right \},
\end{aligned}
\end{align}
where $\text{conv-hull}$ denotes the convex hull of the argument vertices.
\end{assumption}

In the data-driven case, the vertex set $\left \{(A_i,B_i), i=1,\hdots,n \right \}$ can be inferred from data, with its convex hull representing some parameter uncertainty set \cite{petersen2014robust}. That is, instead of identifying a na\"ive certainty equivalence model \cite{aastrom2012introduction}, we can identify an uncertainty set that contains the true parameters of the system. In contrast, in the gain-scheduling case, these vertices can represent the local behavior of $f$ over some grid in the state space, that is,
\begin{align} \label{eq: ABs as jacobians}
    \left[A_i \quad B_i \right]=  \left [ \frac{\partial f}{\partial x} \quad \frac{\partial f}{\partial u} \right ] _{(x,u)=(\bar x^i, \bar u^i)},
\end{align}
where $\left \{(\bar x^i, \bar u^i), i=1,\hdots,n \right \}$ are grid points in some region in the state-input space predetermined by a control engineer/designer.

\subsubsection*{Problem statement} Design a control policy of the form\footnote{In the case of using an approximant lower-order model, $x_k$ can be replaced by the lower dimension version $z_k$, the state of the lower-order model.} $u_k = K x_k+v_k$ that minimizes the cost $J$ in \eqref{eq:costFunction} while stabilizing the system \eqref{eq:stateEquation}, where $v_k$ is a zero mean white noise with covariance $V \succ 0$ and is independent from $w_k$.

Quadratic stability is an important result in the linear control design literature, both for discrete-time systems \cite{amato1998note} and for continuous-time systems \cite{corless1994robust}. In particular, for our subsequent derivations, \cite[Theorem~1]{amato1998note} if $K$ stabilizes $(A_i,B_i),\,i=1,\hdots,n$, then $K$ stabilizes every $(A, B)$ in $\mathbb C$.\footnote{The original result in \cite{amato1998note} is more general, but we pick this restricted case for ease of notation.} Robust and gain-scheduling control guarantees are achieved through satisfying the quadratic stability condition \cite{boyd1993control}. However, Assumption~\ref{assumption:diff inclusion} by itself is not sufficient to satisfy quadratic stability for a new control design. In particular, when $f$ is nonlinear, applying a new control policy (say $K_1 \neq K_0$) may result in distributional shifts in the state-input space compared with the distribution found in data or the grid. These distributional shifts in the state-input space may imply distributional shifts in the parameter space of the approximant difference inclusion model, which in turn weakens Assumption~\ref{assumption:diff inclusion}, the basic and necessary assumption on which these techniques are based. In this work, we seek to dampen these distributional shifts such that Assumption~\ref{assumption:diff inclusion} is held with high validity in the new control design and hence, the quadratic stability condition can be retained.

\section{Background} \label{section: Background}
Robust control is more plausible than the certainty equivalence approach, as ensuring the unknown $(A,B)$ to be within some ambiguity set is generally easier and more likely than identifying an exact model. This is also more plausible in the gain-scheduling case since the difference inclusion \eqref{eq:diff inclusion} is a more flexible and broader model compared with a local difference equation, in encapsulating the behavior of a nonlinear system of the form \eqref{eq:stateEquation}. 

We first seek to approximate the cost by means of the difference inclusion \eqref{eq:diff inclusion}. For each vertex $(A_i,B_i),\,i=1,\hdots,n$, the cost \eqref{eq:costFunction}, up to an additive constant, is
\begin{align*} 
    J_i &= \trace \Big ( \left [ Q + K^\top R K \right ] \sum_{k=0}^\infty \Big \{[A_i+B_iK]^k \times \\
    &\hskip25mm \left [B_i V B_i^\top + W\right][A_i+B_iK]^{k\,\top} \Big \}\Big ),
\end{align*}
or
\begin{equation}\label{eq:cost Controllability type (i)}
\begin{aligned} 
    J_i &= \trace \left (\left [ Q + K^\top R K \right ] \Sigma_i  \right ),
\end{aligned}
\end{equation}
where $\Sigma_i $ is the controllability-type Gramian (the steady-state covariance of the state, assuming the state behaves according to the vertex model), given by
\begin{align}
    \Sigma_i  &= \lim_{k \to \infty}\E \left \{x_k x_k^\top \right \}  =\sum_{k=0}^\infty \Big \{ \left [ A_i+B_iK\right ]^k \times \nonumber
    \\ &\hskip25mm
    \left [B_i V B_i^\top + W\right] \left [ A_i+B_iK\right ]^{k\,\top} \Big \}, \label{eq:Sigma as state covariance}
\end{align}
and is also the solution of the Lyapunov equation
\begin{align}
    \Sigma_i = \left [ A_i+B_iK\right ] \Sigma_i \left [ A_i+B_iK\right ]^\top + B_iV B_i^\top + W. \label{eq:Controllability Lyapunov}
\end{align}
The controllability-type Gramian is positive definite, $\Sigma_i \succ 0$, since $V \succ 0$ and $(A_i,B_i)$ is assumed stabilizable.

The above is the cost and the stability guarantee (a controller $K$ that satisfies the Lyapunov equation \eqref{eq:Controllability Lyapunov}) if the state behaves according to the vertex model $(A_i, B_i)$. However, if the difference inclusion model \eqref{eq:diff inclusion} is to be adopted, the quadratic stability condition is achieved by a controller $K$ that guarantees the existence of one $\Sigma \succ 0$ such that
\begin{align*}
    \Sigma &\succeq \left [ A_i+B_iK\right ] \Sigma \left [ A_i+B_iK\right ]^\top + B_iV B_i^\top + W,
\end{align*}
for $i=1,2,\hdots,n.$ The above nonlinear matrix inequality is equivalent to the LMI condition \cite{boyd1993control}, with the change of variables $L = K \Sigma$,
\begin{align*}
    \begin{bmatrix}
        \Sigma-B_i V B_i^\top-W & A_i \Sigma + B_i L \\
        \Sigma A_i^\top + L^\top B_i^\top & \Sigma
    \end{bmatrix} \succeq 0, \,i=1,2,\hdots,n.
\end{align*}

This $\Sigma$ can also be used instead of $\Sigma_i$ in \eqref{eq:cost Controllability type (i)}, as a description of the collective performance of the difference inclusion \eqref{eq:diff inclusion}.
\begin{align*}
    \tilde J = \trace \left (  Q\Sigma  \right ) + \trace \left ( R K \Sigma  K^\top \right).
\end{align*}
To achieve an objective function that is linear in the decision variables $\Sigma$ and $L$, we define the extra variable $Z_0$, such that $Z_0 \succeq K \Sigma  K^\top$. This inequality can be equivalently described by the LMI:
\begin{align*}
    \begin{bmatrix}
        Z_0 & K \Sigma\\
        \Sigma K^\top & \Sigma
    \end{bmatrix} \succeq 0, \text{ or, }
    \begin{bmatrix}
        Z_0 & L\\
        L^\top & \Sigma
    \end{bmatrix} \succeq 0.
\end{align*}

The robust (and gain-scheduling) control in the controllability-type parametrization can now be described by the following problem.
\vskip 2mm
\textbf{Robust LQR control:}
\begin{equation} \label{eq:Robust LQR problem controllability}
    \begin{aligned}
         &\min_{\Sigma ,\,L,\,Z_0} \trace \left (  Q\Sigma  \right ) + \trace \left ( R Z_0 \right) \\
        &\text{s.t. } \Sigma \succ 0, \quad
        \begin{bmatrix}
            Z_0 & L\\
            L^\top & \Sigma 
        \end{bmatrix}\succeq 0, \\
        &
        \begin{bmatrix}
        \Sigma-B_i V B_i^\top-W & A_i \Sigma + B_i L \\
        \Sigma A_i^\top + L^\top B_i^\top & \Sigma
    \end{bmatrix} \succeq 0, \,i=1,\hdots,n.
    \end{aligned}
\end{equation}

The solution controller can then be recovered from the solution $\Sigma_\star,L_\star$ by $K_\star=L_\star \Sigma_\star^{-1}$.

This parametrization is more convenient for our derivation (compared to the more common observability-type parametrization) because, as expressed by \eqref{eq:Sigma as state covariance}, the controllability-type Gramian can be related to the steady-state covariance of the state. Hence, it has a statistical interpretation and can be used in comparison to the learning data when enforcing data consistency.

\section{Dampened parameter distributional shifts}
The quadratic stability condition holds when the underlying true system \eqref{eq:stateEquation} is captured by the difference inclusion \eqref{eq:diff inclusion} under any condition and any controller $K_0$. This might be possible to achieve when the underlying system is linear and possibly time-varying with a range of parameters (and hence a convex hull) that can be predetermined. However, Assumption~\ref{assumption:diff inclusion} becomes problematic if the underlying system is nonlinear. Specifically, if a newly designed (possibly robust according to \eqref{eq:Robust LQR problem controllability}) state feedback gain $K$ is applied, the resulting state-input distribution of the new closed-loop system is not necessarily the same as that used in the assumption for the robust control case or the same as the grid in the gain-scheduling case under the controller $K_0$. This may imply that the difference inclusion \ref{eq:diff inclusion} is no longer capturing the behavior of the underlying system, and therefore, the control design process itself, even if ``robust'' as presented in the above program, may invalidate the basic premises on which quadratic stability is built on.

We show next how to mitigate the above problem by ``dampening'' the distributional shifts by adopting the data-conforming concept introduced in \cite{ramadan2024data}.

\subsection{Data conforming in the state-input distribution}
We designate the new design and data (grid) state-input distributions\footnote{The notation $ \mathcal{N}(\mu_0, \Sigma_0)$ denotes a Gaussian density of mean $\mu$ and covariance $\Sigma$.} by $\mathcal{N}_{des} =  \mathcal{N}(0, \Gamma_{des})$ and $\mathcal{N}_{data} =  \mathcal{N}(0, \Gamma_{data})$, respectively\footnote{The state-input distributions are assumed Gaussian or accurately approximated by their first two moments.}. Therefore, the covariance matrices $\Gamma_{des},\Gamma_{data}$ fully characterize these distributions.

The data covariance matrix satisfies
\begin{align}
\begin{aligned} \label{eq:Gamma data def}
    \Gamma_{data}&\approx \frac{1}{n+1} 
    \begin{bmatrix}
        x_{k,data}\\
        u_{k,data}
    \end{bmatrix}
    \begin{bmatrix}
        x_{k,data}\\
        u_{k,data}
    \end{bmatrix} ^\top = 
    \begin{bmatrix}
         \Sigma_{data} &  H_{data}\\
         H_{data}^\top &  M_{data}
    \end{bmatrix},
\end{aligned}
\end{align}
where $H_{data} = (n+1)^{-1} \sum_{k=1}^n x_{k,data} u_{k,data}^\top$ and $M_{data} = (n+1)^{-1} \sum_{k=1}^n u_{k,data} u_{k,data}^\top$. On the other hand, the design covariance matrix satisfies
\begin{align} \label{eq:design covariance}
    \Gamma_{des} &= 
    \begin{bmatrix}
        \lim_{k \to \infty} \E x_k x_k^\top & \lim_{k \to \infty} \E x_k u_k^\top \\
        \lim_{k \to \infty} \E  u_k x_k^\top & \lim_{k \to \infty} \E  u_k u_k^\top
    \end{bmatrix}\nonumber\\
    &= 
    \begin{bmatrix}
        \Sigma & \Sigma K^\top \\
        K\Sigma & K\Sigma K^\top + V
    \end{bmatrix},
\end{align}
since $u_k = K x_k + v_k$. Here, $\Sigma$ is as defined in \eqref{eq:Sigma as state covariance}. 

 We use Jeffreys divergence between the densities $ \mathcal{N}_{des}$ and $ \mathcal{N}_{data}$. This divergence term reduces to an expression in terms of the covariances of these densities, which under some relaxation can be posed as an affine regularization term and LMI constraints in $\Sigma$ and $L$ (from the change of variables $L=K \Sigma$).

Following the same derivation in \cite[Lemma~1 and the subsequent discussion]{ramadan2024data}, the Jeffreys divergence regularization term is given by
\begin{align}
    F(\Gamma_{des}) = \trace \left ( \Gamma_{data}^{-1}\Gamma_{des} + \Gamma_{data}\Gamma_{des}^{-1}\right). \label{eq:F}
\end{align}
The function $F$ possesses favorable properties, making it convenient as a regularization term: $\Gamma_{des} = \Gamma_{data}$ is the global minimizer of $F(\Gamma_{des})$ and $F$ is convex in $\Gamma_{des}$. This is discussed in more details in \cite{ramadan2024data}, together with showing how the Jeffreys divergence results from linearizing the Kullback-Leibler divergence between Gaussian densities. This connection between the Jeffreys divergence and the Kullback-Liebler divergence is important, since the latter is widely acknowledged and used in the contexts of consistency to learning data and the exploration versus exploitation trade-offs \cite{schulman2017proximal}.

For the first term of $F$ in \eqref{eq:F}, $\trace \left(\Gamma_{data}^{-1}\Gamma_{des} \right)$, we can introduce the extra variable $Z_1$, such that $Z_1 \succeq \Gamma_{des}$, or equivalently
\begin{align*} 
    Z_1 \succeq \Gamma_{des} &= \begin{bmatrix}
        \Sigma & \Sigma K^\top \\
        K\Sigma & K\Sigma K^\top + V
    \end{bmatrix} \\
    &=
    \begin{bmatrix}
        \Sigma\\
        K \Sigma
    \end{bmatrix} \Sigma^{-1}
    \begin{bmatrix}
        \Sigma & \Sigma K^\top
    \end{bmatrix}+
    \mathcal{V},
\end{align*}
or as an LMI in $\Sigma$ and $L$
\begin{align}\label{eq:Z_1 LMI}
\begin{bmatrix}
        Z_1 - 
        \mathcal{V}
    & \begin{bmatrix}
        \Sigma\\
        L
    \end{bmatrix}\\
    \begin{bmatrix}
        \Sigma & L^\top
    \end{bmatrix} & \Sigma
\end{bmatrix} \succeq 0,
\end{align}
where $\mathcal{V}= \text{block-diag}(0_{r_x \times r_x},V)$. In the second term of \eqref{eq:F}, $\Gamma_{des}^{-1}$, as an inverse of a partitioned matrix \cite{horn2012matrix}, is given by
\begin{align*}
\Gamma_{des}^{-1} = 
\begin{bmatrix}
        \Sigma^{-1}+K^\top V^{-1} K & -K^\top V^{-1} \\
        -V^{-1} K & V^{-1}
\end{bmatrix},
\end{align*}
which exists and is positive definite, since $\Gamma_{des} \succ 0$. Using the subsequent discussions and derivations in \cite{ramadan2024data}, the above term can be approximated, up to an additive constant (in $\Sigma$ and $L$), by a linear term in the objective, subject to LMI constraints. The resulting Data-conforming robust (gain-scheduling) LQR control is given as follows.

\vskip 2mm
\textbf{Data-conforming robust LQR control:}
\begin{equation} \label{eq:state-input-data-conforming Robust LQR problem}
    \begin{aligned}
         &\min_{\Sigma ,\,L,\,Z_0,\,Z_1,\,Z_2,\,Z_3} \trace \left (  Q\Sigma  \right ) + \trace \left ( R Z_0 \right) + \\
         &\gamma \Big \{ \trace \left ( \Gamma_{data}^{-1} Z_1 \right) + \trace \left ( V^{-1} Z_2 \right) + \trace \left ( \Sigma_{data} Z_3 \right) \Big \} \\
        &\text{s.t. } \Sigma \succ 0, \quad
        \begin{bmatrix}
            Z_0 & L\\
            L^\top & \Sigma 
        \end{bmatrix}\succeq 0,\\
        &\begin{bmatrix}
        \Sigma-B_i V B_i^\top-W & A_i \Sigma + B_i L \\
        \Sigma A_i^\top + L^\top B_i^\top & \Sigma
    \end{bmatrix} \succeq 0, \\
    &i=1,\hdots,n,\\
    &\begin{bmatrix}
        Z_1 - 
        \mathcal{V}
    & \begin{bmatrix}
        \Sigma\\
        L
    \end{bmatrix}\\
    \begin{bmatrix}
        \Sigma & L^\top
    \end{bmatrix} & \Sigma
\end{bmatrix} \succeq 0, \quad
\begin{bmatrix}
        Z_3 & I \\
    I & \Sigma
\end{bmatrix} \succeq 0, \\
&
\begin{bmatrix}
        Z_2 & L- H_{data}^\top \Sigma_{data}^{-1} \Sigma \\ 
        \left[L- H_{data}^\top \Sigma_{data}^{-1} \Sigma\right]^\top & \Sigma
    \end{bmatrix} \succeq 0,
    \end{aligned}
\end{equation}
and the solution control gain can be recovered from the optimal values $\Sigma_\star$ and $L_\star$ by evaluating $K_\star = L_\star \Sigma_\star^{-1}$.

The following result shows that even though the covariance matrix $\Sigma_\star$ is not necessarily the actual covariance matrix of the true underlying system after applying the control gain $K_\star$, it is nevertheless an upper bound to the actual covariance.
\begin{corollary}
    In the solution $(\Sigma_\star, K_\star)$ of \eqref{eq:state-input-data-conforming Robust LQR problem}, $\Sigma_\star$ is an upper bound of all the $\Sigma_i$s defined with $K=K_\star$ in \eqref{eq:Controllability Lyapunov}.
\end{corollary}
\begin{proof}
The third LMI in \eqref{eq:state-input-data-conforming Robust LQR problem} is equivalent to \cite{boyd1994linear}
\begin{align*}
    \Sigma \succeq \left [ A_i+B_iK\right ] \Sigma \left [ A_i+B_iK\right ]^\top + B_iV B_i^\top, \,i=1,\hdots,n.
\end{align*}
One can show that the minimal covariances satisfying each of the above inequalities is, in fact, the one that satisfies the equality (active constraint). That is $\Sigma_\star \succeq \Sigma_i$, for all $i$. This can be seen from the optimality conditions of the problem of minimizing $\trace(\Sigma)$ subject to the above LMIs, since when $K = K_\star$, by the primal feasibility,
\begin{align*}
    \Sigma_\star \succeq \left [ A_i+B_iK_\star\right ] \Sigma_\star \left [ A_i+B_iK_\star\right ]^\top + B_iV B_i^\top,
\end{align*}
while by definition of $\Sigma_i$ in \eqref{eq:Controllability Lyapunov}, when $K=K_\star$,
\begin{align*}
    \Sigma_i = \left [ A_i+B_iK_\star\right ] \Sigma_i \left [ A_i+B_iK_\star\right ]^\top + B_iV B_i^\top.
\end{align*}
\end{proof}

The above result ensures that the state-input distribution of the actual system is ``concentrated inside'' that of the design ($\Gamma_{true} \preceq \Gamma_{des}$). Therefore, if \eqref{eq:state-input-data-conforming Robust LQR problem} manages to achieve $\Gamma_{des} \approx \Gamma_{data}$, the true distribution is concentrated inside the data distribution, and consistency (data-conformity) is achieved.

\begin{corollary}
    If the original robust control problem \eqref{eq:Robust LQR problem controllability} is feasible, then the data-conforming robust control \eqref{eq:state-input-data-conforming Robust LQR problem} is feasible as well.
\end{corollary}
\begin{proof}
    The extra LMI conditions in \eqref{eq:state-input-data-conforming Robust LQR problem} are merely a consequence of the regularization terms added to the cost of the original robust control problem and do not alter its feasibility.
\end{proof}

\begin{remark} 
One can relax the zero-mean condition in the state-input distributions, that is, $\mu_{des},\mu_{data} \neq 0$. The cost \eqref{eq:costFunction} can be shown to be quadratic in the variables $\bar x$ and $\bar u$, where $\mu_{des} = \left [\bar x^\top,\, \bar u^\top \right ]^\top$. The variables $\bar x$ and $\bar u$ also appear in the linear constraint given by the certainty equivalence dynamics
\begin{align*}
    \bar x = (A + B K) \bar x + B \bar u.
\end{align*}
Therefore, we can construct a two-stage optimization problem. First, we solve \eqref{eq:state-input-data-conforming Robust LQR problem} for the optimal gain $K_\star$ and covariance $\Sigma_\star$. Second, with the cost being quadratic in $\mu_{des}$, we solve for the optimal $\bar u$. This process can be repeated iteratively and incrementally with a carefully chosen exploration and exploitation balance.
\end{remark}

The proposed formulation \eqref{eq:state-input-data-conforming Robust LQR problem} is a convex SDP with an affine cost and LMI constraints. Hence, this guarantees a level of scalability to handle problems with high state-input dimensions. In the next section we instead present a simple example to showcase the failures of regular robust control in stabilizing nonlinear systems and further explain how the data-confroming concept can improve the validity of the quadratic stability condition.

\section{Numerical simulations} \label{section: Numerical}
 \label{section: Numerical}

Consider the following dynamic system:\footnote{The results of this section can be reproduced by our open-source \textsc{Julia} code found at \href{https://github.com/msramada/robust-less-dist-shifts}{https://github.com/msramada/robust-less-dist-shifts}.} 
\begin{equation}
\begin{aligned} \label{eq:example dynamic system}
    x_{k+1} &= 
    \begin{bmatrix}
        .98& .1\\
        0& .95
    \end{bmatrix}x_k + 
    \begin{pmatrix}
        .25 x_{2,k}^2\\
        0
    \end{pmatrix} + \\
    &\hskip 15mm
    \begin{bmatrix}
        0\\
        0.1 + .25 \tanh{x_{1,k}}
    \end{bmatrix}u_k + w_k,
\end{aligned}
\end{equation}
where $w_k$ is normally distributed, white, of zero mean, and with covariance $W=diag(\left [ .2,\, .1\right])$. We choose this system as it contains a state-input nonlinearity, introducing a coupling in the state-input space. Also this system contains the term $.25 x_{2,k}^2$ which can have a small effect when $x_{2,k}$ is small (not effectively explored) but can have a significant and destabilizing effect if a new controller is applied that is aggressive in the $x_{2,k}$ direction.

According to Remark~\ref{remark: 2 cases covered}, if the above system is fully or partially unknown, a difference inclusion of the form \eqref{assumption:diff inclusion} can be identified from the data. This difference inclusion can also model the above system if it is known and a gain-scheduling control is to be designed. That is, the vertices $(A_i, B_i)$ constructing this inclusion are either inferred from data or, in the gain-scheduling case, given by the Jacobians,
\begin{equation}\label{eq:samples2vertex example}
\begin{aligned} 
    A_i &= \left. \frac{\partial f (x, u,w)}{\partial x} \right |_{(x,u,w)=(\bar x^i, \bar u^i, 0)} \\
    &= 
    \begin{bmatrix}
        .98& .1+.5 \bar x_{2}^i\\
        .25 (1-\tanh^2{\bar x_{1}^i})\bar u^i& .95
    \end{bmatrix}, \\
    B_i &= \left. \frac{\partial f (x, u,w)}{\partial u} \right |_{(x,u,w)=(\bar x^i, \bar u^i, 0)} \\
    &=
    \begin{bmatrix}
        0\\
        0.1 + .25 \tanh{\bar x_{1}^i}
    \end{bmatrix},\quad i=1,\hdots,n,
\end{aligned}
\end{equation}
where $x^i=[x_1^i, x_2^i]^\top$ and $\bar u_i$ for $i=1,\hdots,n$ are the grid points used to approximate the local behavior of the system in the region they are sampled from, as in \eqref{eq: ABs as jacobians}. 

With $Q=diag([1,\,.5])$, $R=1$, and $V=.05$, we implement the following steps: (i) we sample $n=500$ of the $\bar x^i=[\bar x_1^i, \bar x_2^i]^\top$ and $\bar u_i$ independently according to a Gaussian density of zero mean and covariance $\Gamma_{data}$ compose of $\Sigma_{data}=.5\mathbb I_{2 \times 2}$ and $M_{data}=.5$ (here $H_{data}=0$), respectively, then evaluate the corresponding vertex points $(A_i, B_i)$ according to \eqref{eq:samples2vertex example}. Using the quickull algorithm \cite{barber1996quickhull} in \textsc{Julia}, Quickhull.jl, we find the vertices representing the convex hull containing all other vertices (they are $39$ in this step), (ii) we evaluate the quadratically stable LQR control $K_{robust}$ according to \eqref{eq:Robust LQR problem controllability}. Then (iii) we apply this control in feedback for the system \eqref{eq:example dynamic system} for $500$ time steps; and (iv) for the resulting state-input trajectory $\{x_k^{robust},u_k^{robust}\}_{k=1}^{500}$, we evaluate the corresponding state-space matrices as in \eqref{eq:samples2vertex example}, replacing $(\bar x^i,\bar u^i)$ with $(x_k^{robust},u_k^{robust})$, to get $(A_k^{robust},B_k^{robust}),\,k=1,\hdots,500$. To solve these SDPs we use an interior point conic optimization solver in \textsc{Julia}, Clarabel.jl \cite{Clarabel_2024}.

Similarly, we apply the three steps (ii)--(iv) to evaluate the data conforming control $K_{DC}$ according to \eqref{eq:state-input-data-conforming Robust LQR problem}, with $\gamma = 10$, then use this control as in step (iii) to sample a $500$-timestep state-input trajectory and evaluate the corresponding matrices $(A_k^{DC},B_k^{DC})$. For the purpose of comparison, we also include the LQR controller $K_{LQR}$ of the local linearization about the origin, and similarly evaluate $(A^{LQR}_k,B_k^{LQR})$.

\begin{figure*}
\centering 
\includegraphics[width=5.5in,height=2.2in]{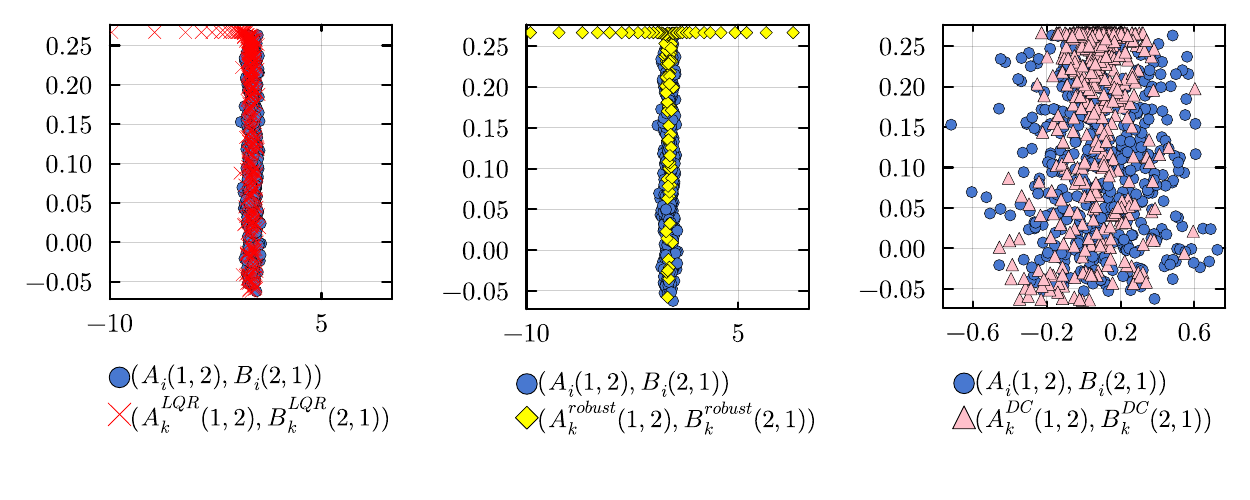} 
\caption{Blue circles are the parameters $(A_i(1,2),B_i(2,1))$ over the grid points, given by \eqref{eq:samples2vertex example}, and used in the control design process. Red crosses, yellow diamonds, and pink triangles correspond to the parameters (of the same indices) of the local approximation matrices over the trajectories achieved after the application of $K_{LQR}$, $K_{robust}$, and $K_{DC}$, respectively.}\label{fig:distShifts}
\end{figure*}

Figure~\ref{fig:distShifts} illustrates the resulting distributional shifts after applying each of the controllers $K_{LQR}$, $K_{robust}$, and $K_{DC}$ (left to right, in the same order as the subplots). The x-axis of all the subplots corresponds to the $(1,2)$ entry of the state matrices $A_k^{LQR}, A_k^{robust}, A_k^{DC}$, while the y-axis corresponds to the $(2,1)$ entry of the input matrices $B_k^{LQR},B_k^{robust},B_k^{DC}$. Notice the ``leakage'' in the parameters in the $K_{LQR}$ and $K_{robust}$ cases outside of the grid distributions, which eventually results in divergence and instability of the closed-loop system.

We repeat the above four steps, from the sampling step (i) to the evaluation of the state-input trajectories of the three controllers $K_{DC}, K_{robust}, K_{LQR}$, for 1,000 times. We then count the number of stable simulations resulting from each control gain based on the boundedness of the state under some norm (in particular, we use the infinity norm $\lVert x_k \rVert_{\infty} \geq 100$ as an instability check), pointwise in time for $500$ time steps. We report the percentages of the stable simulations in Table~1.

The instability after applying $K_{LQR}$ is due to the assumption that the system will be very close to the origin, which is turned out to be wrong. $K_{robust}$ shows some improved results since it is built on the modeling flexibility of the difference inclusion over a grid. However, the state-input data, after applying $K_{robust}$, can be distributed differently than the grid points $(x_i,u_i)$, thus, it is no longer described by the difference inclusion used in its design, invalidating the premise of the quadratic stability condition. In contrast, our data-conforming $K_{DC}$ enforces the consistency of the closed-loop system state-input data with the grid data, dampening parameter distributional shifts and solidifying the assumption of the difference inclusion as a model.

\begin{table}
\begin{center}
\begin{tabular}{ |c|c|c| } 
 \hline
 \multicolumn{3}{|c|}{Table 1: Percentages of stable simulations (out of 1,000 simulations)} \\
 \hline
 LQR (about origin) & Robust \eqref{eq:Robust LQR problem controllability} & Data-conforming robust \eqref{eq:state-input-data-conforming Robust LQR problem}\\
  \hline
  0.0\% & 64.9\% & 94.8\% \\
 \hline
\end{tabular}
\end{center}
\end{table}

\section{Conclusion} \label{section: Conclusion}
This paper seeks to mitigate a problem inherent in data-driven robust control and gain-scheduling approaches when applied to nonlinear systems. In particular, the distributional shifts in the parameter space of the approximant models when new control design is implemented. That is, the parameter distribution of the approximant model used in the design might be very different from that after the implementation of the new design in closed-loop. This in effect can invalidate the basic premise on which quadratic stability and therefore robust control are built on. To mitigate this problem, our proposed data-conforming approach enforce distributional consistency by dampening any shifts in the state-input space, and hence dampening any shifts in the parameters of the approximant model. Our methods are solutions of semi-definite programs of affine costs and linear matrix inequalities, making them computationally efficient and scalable up to systems with hundreds in state-input dimensions.

Our future work is directed toward expanding the data-conforming control framework to modern data-driven optimal control design techniques. We are also investigating the possibility of developing a data-conforming policy gradient that, in contrast to standard policy gradient methods, can dampen distributional shifts in the state-input spaces during learning.

\bibliographystyle{IEEEtranS}
\bibliography{References}

\vspace{0.1cm}
\begin{flushright}
	\scriptsize \framebox{\parbox{2.5in}{Government License: The
			submitted manuscript has been created by UChicago Argonne,
			LLC, Operator of Argonne National Laboratory (``Argonne").
			Argonne, a U.S. Department of Energy Office of Science
			laboratory, is operated under Contract
			No. DE-AC02-06CH11357.  The U.S. Government retains for
			itself, and others acting on its behalf, a paid-up
			nonexclusive, irrevocable worldwide license in said
			article to reproduce, prepare derivative works, distribute
			copies to the public, and perform publicly and display
			publicly, by or on behalf of the Government. The Department of Energy will provide public access to these results of federally sponsored research in accordance with the DOE Public Access Plan. http://energy.gov/downloads/doe-public-access-plan. }}
	\normalsize
\end{flushright}	
\end{document}